\documentclass[showpacs,showkeys,12pt,aps]{revtex4}
\usepackage{graphicx}
\usepackage{hyperref}
\usepackage{latexsym}
\usepackage{amssymb}
\usepackage{color}

\begin{document}

\title{Buchdahl-Vaidya-Tikekar model for stellar interior in pure Lovelock gravity - II} 

\author{Alfred Molina}
  \affiliation{Departament de F\'\i sica Qu\`antica i Astrof\'{\i}sica,\\ Institut de Ci\'encies del Cosmos.  Universitat de Barcelona, Spain}
  \email{alfred.molina.ub.edu}

 \author{Naresh Dadhich}
\affiliation{Centre for Theoretical Physics, Jamia Millia, Islamia, New Delhi, 110025, India}
 \affiliation{Inter-University Center for Astronomy and Astrophysics,
 Post Bag 4 Pune 411 007, India}
 \email{nkd@iucaa.in}

 \author{Avas Khugaev}
 \affiliation{Institute of Nuclear Physics,
 Tashkent, 100214, Uzbekistan}
 \email{avaskhugaev@mail.ru }
\date{\today}

\begin{abstract}
For a given Lovelock order $N$, it turns out that static fluid solutions of the pure Lovelock equation for a star interior have the universal behavior in all $n\geq 2N+2$ dimensions relative to an appropriately defined variable and the Vaidya-Tikekar parameter $K$, indicating deviation from sphericity of $3$-space geometry. We employ the Buchdahl metric ansatz which encompasses almost all the known physically acceptable models including in particular the Vaidya-Tikekar and Finch-Skea. Further for a given star radius, the constant density star, always  described by the Schwarzschild interior solution, defines the most compact state of distribution while the other end is marked by the Finch-Skea model, and all the other physically tenable models lie in between these two limiting distributions.   
\end{abstract}
\maketitle

\section{Introduction}

A model for a star interior is an important astrophysical question and hence it is pertinent  to construct relativistic models as solutions of gravitational equation. For constructing static perfect fluid interiors for compact objects, we have two metric functions to determine while there is only one pressure isotropy equation for determining them. We need therefore to prescribe either a metric ansatz for one of the functions or an equation of state relating pressure and density or a fall off behavior for density or pressure. In Ref. \cite{kdm} (henceforth to be referred as paper I) we had considered a fairly general Buchdahl-Vaidya-Tikekar metric (BVT) \cite{Buchdahl, Vaidya} ansatz covering almost all physically tenable known models and had shown that the pressure isotropy equation bore the same Gauss form in all dimensions $\geq 4$. By exploiting this property we established that a $4$-dimensional Einstein solution for a given value of the Vaidya-Tikekar parameter $K_4$, indicating deviation from sphericity of $3$-space geometry, could be lifted to a higher $n$-dimensional solution with $K$ being replaced by $K_n=(K_4 -n +4)/(n-3)$. That is, a static fluid interior solution in the usual $4$ dimension could be taken over to a higher dimensional solution.  

This is an interesting property of fluid solutions in Einstein gravity and so the natural question that arises is that, is this also carried over to Lovelock theory which is quintessentially higher dimensional ? Interestingly the answer is yes, but it singles out pure Lovelock from general Lovelock theory. By pure Lovelock we mean the action and consequently the equation of motion has only one $N$-th order term with no sum over lower orders. That is, a $(2N+2)$-dimensional solution with a given $K_{2N+2}$ can be taken over to higher $n$-dimensional solution with $K_n=(K_{2N+2}-n+2N+2)/(n-2N-1)$. This is yet another property that singles out pure Lovelock gravity from all others. In addition to the Vaidya-Tikekar (VT) \cite{Vaidya}, the Buchdahl ansatz \cite{Buchdahl} also includes in the limit, the interesting Finch-Skea ansatz (FS) \cite{FS} and so we also obtain the pure Lovelock version of the FS solution. In this case the isotropy equation is not in the Gauss but instead in the Bessel form which for a given $N$ is the same for all dimensions. 

Thus there is universality in behavior of star interiors corresponding to both BVT as well as FS models relative to a properly defined variable and also the parameter $K$ for the former. For the latter, for a given $N$ the solution is the same in all dimensions. This is a very important property of fluid solutions making the star interior in the pure Lovelock theory. The compactness hierarchy goes as the constant density star defining the limiting upper limit while FS holding the other end and the Buchdahl including all other physically reasonable models lying between these two limits. For a given radius for star in critical $n=2N+2$, mass is maximum for the Einstein $N=1$ bearing out the fact that pure Lovelock gravity becomes weaker with $N$. We shall therefore study both VT and FS models and their physical properties for pure Lovelock gravity. This is what would be the main subject matter of this paper, part II of paper I. 

There is a fairly large body of work on higher dimensional star which was comprehensively reviewed in paper I. Instead of repeating that again here we would like to direct the reader to paper I \cite{kdm} for detailed references. It should however be admitted that higher dimensional fluid models are more for exploring and probing the gravitational dynamics rather than their direct physical and astrophysical applications. Of course there always remains a little window open for emergence of a unified field theory involving higher dimensions where it may find some relevance. 

The paper is organized as follows: In the next section we recall the pure Lovelock gravitational equation and set it for a perfect fluid distribution. In Sec III, we specialize to static spherically symmetric with the Buchdahl ansatz and find solutions for VT and FS ansatzs, which  is followed by matching of interior and exterior solution. The physically properties are discussed in Sec V and we conclude with a discussion.    

\section{Lovelock gravity} 

There is a natural generalization of Einstein action to  Lovelock action which is a homogeneous polynomial in Riemann curvature with Einstein being the linear order and the quadratic is Gauss-Bonnet (GB). It has the remarkable property that on variation it still gives the second order quasi-linear equation which is its distinguishing feature. Note that for pure Lovelock with only one $N$th order term in the action without sum over lower orders, it is   pure divergence -- topological in $n=2N$, and in $n=2N+1$ it has the same behavior of non-existence of non-trivial vacuum solution relative to Lovelock analogue of Riemann tensor \cite{dad2008, cam-dad} as Einstein has in three dimension; i.e. gravity is kinematic. As for Einstein, pure Lovelock gravity turns dynamic for dimensions $\geq 2N+2$. Clearly Lovelock is therefore a  quintessentially higher dimensional gravitational theory for $n\geq 2N+1$. 
 
If we  introduce a set of $(2N,2N)$-rank tensors \cite{Kastor} product of $N$ Riemann tensors, completely antisymmetric, both in its upper and lower indices,
\begin{equation}
^{(N)}\!I\!\!R^{b_1 b_2 \cdots b_{2N}}_{a_1 a_2 \cdots a_{2N}}= R^{[b_1 b_2}_{\quad \quad [a_1 a_2}\cdots R^{b_{2N-1} b_{2N}]}_{\qquad \qquad a_{2N-1} a_{2N}]}~.
\label{kastensor}
\end{equation}
With all indices lowered, this tensor is also symmetric under the exchange of both groups of indices, $a_i\leftrightarrow b_i$. 
In terms of these new objects we can now write
\begin{equation}
L_N= {}^{(N)}\!I\!\!R \qquad \text{and} \qquad
{}^{(N)}\mathbf{G}^\mu_{\ \nu}=N\,\,{}^{(N)}\!I\!\!R^\mu_{\ \nu}- \frac12 {}^{(N)}\!I\!\!R\,\delta^\mu_{\ \nu}
\end{equation} 

For pure Lovelock of order $N$ we write  
\begin{equation} 
S=\int d^nx\sqrt{-g}{L}_{N}+S_{\rm matter},\label{action}
\end{equation}

Now the pure Lovelock gravitational equation for $n\geq 2N+1$ has the usual form 
\begin{equation}
{}^{(N)}\mathbf{G}^\mu_{\ \nu} =  T^\mu_\nu,\quad \mu,\nu=1\dots n  \label{leq}
\end{equation}

We are going to solve this equation for a static spherically  symmetric fluid distribution with $T^\mu_\nu = diag(-\rho, p, p, ..., p)$ as a model for star interior in hydrostatic equilibrium.  

\section{Buchdahl ansatz} 

Let us begin with the general static spherically symmetric metric in  $n$--dimensional spacetime,
\begin{eqnarray}
ds^2=-f_2(r)^2 dt^2+\frac{1}{1-f_1(r)}dr^{2}+r^{2}d\Omega^{2}_{n-2}\,.
\label{eqn_1}
\end{eqnarray}
Substituting this metric in $N$th order pure Lovelock equation Eq. (\ref{leq}), we obtain
\begin{equation}
\rho=\frac12\frac{(n-2)!}{(n-2N-1)!}\left(\frac{f_1(r)}{r^2}\right)^N\left( Nr\frac{f'_1(r)}{f_1(r)}+n-2N-1\right)
\label{eqn_3}
\end{equation}

\begin{equation}
p=\frac{(n-2)!}{(n-2N-1)!}\left(\frac{f_1(r)}{r^2}\right)^N\left( Nr\frac{1-f_1(r)}{f_1(r)}\frac{f'_2(r)}{f_2(r)}-\frac12 (n-2N-1)\right)
\label{eqn_4}
\end{equation}
and the pressure isotropy equation is given by  

\begin{eqnarray}
p_\theta-p_r &=&\frac{N(n-3)!}{(n-2N-1)!}\left(\frac{f_1(r)}{r^2}\right)^{N-1}\frac{1}{f_2(r)}\Bigg[ (1-f_1(r))f''_2(r)-\nonumber\\ 
&& \frac{f'_2(r)}{r}\left\{(n-2)(1-f_1(r))+\frac12 rf'_1(r)\left(n-2-\frac{n-3}{f_1(r)}\right)\right\}- \nonumber\\ &&
\frac{n-2N-1}{2}\left(\frac{f'_1(r)}{f_1(r)}-\frac2r\right)\left(f'_2(r)(1-f_1(r))+\frac{f_1(r)f_2(r)}{r}\right)\Bigg]\label{eqn_2} = 0 
\end{eqnarray}
where a prime indicates derivative relative to $r$. It may be noted that these are general expressions for density, pressure and pressure isotropy for any Lovelock order $N$ which have perhaps not been reported earlier anywhere. These would therefore be useful for all future considerations.  

Before we go any further let's rule out the critical odd $n=2N+1$ dimension case from further discussion. For a bound distribution describing interior of a compact object, the boundary is defined by $p=0$ which requires from equation (\ref{eqn_4}), $f'_2(r)=0$ on the boundary. This conflicts with the matching with exterior vacuum solution. There cannot therefore exist bound distribution in the critical odd $n=2N+1$ dimensions \cite{dhm, sudan}. We shall henceforth only consider $n\geq 2N+2$. 

Note that we have only one equation  (\ref{eqn_2}) to determine the two unknown metric functions $f_1(r)$ and $f_2(r)$ while the other equations (\ref{eqn_3} - \ref{eqn_4}) define the density and the pressure. Hence it is imperative either to have an ansatz specifying one of the metric functions or an equation of state relating density and pressure or a fall off behaviour for density. 

We resort to a fairly general ansatz due to Buchdahl \cite{Buchdahl} prescribing the metric function $f_1(r)$ as 
\begin{equation}
f_1(r) = \frac{Ar^2}{1+Cr^2}\label{eq_B}
\end{equation}
with $A > C > 0$. Vaidya and Tikekar (VT) further particularized \cite{Vaidya} it by writing $C=K\alpha, A=(1+K)\alpha$, and wrote 
\begin{eqnarray}
f_1(r)=\frac{(1+K)\alpha r^{2}}{1+K\alpha r^{2}}\,
\label{eqn_5}
\end{eqnarray}
where $\alpha = R^{-2}$ ($K$ here is $-K$ in \cite{Vaidya}). They had given this parameter an interesting geometric meaning as deviation from sphericity of $3$-space geometry. It may also be noted that this parameter is required to be positive for density to be monotonically decreasing outwards from the center of distribution \cite{kdm}. It is interesting that this is how $3$-space geometry is related to density evolution of fluid. 

Another interesting ansatz is due to Finch and Skea \cite{FS} which is the limiting case of the Buchdahl ansatz when $A=C$, and then 
\begin{eqnarray}
f_1(r) = \frac{Cr^2}{1+Cr^2}.
\end{eqnarray}

It should be stated that though there are solutions which are lying outside  the Buchdahl ansatz but they all seem to suffer from one or the other unphysical feature like density increasing outwards \cite{exactsol}. So Buchdahl ansatz covers all the physically tenable star models. We would therefore like to employ the ansatz for studying star interiors in pure Lovelock gravity. In particular we would like to obtain solutions for star interior for the two ansatzs: Vaidya-Tikekar (VT) and Finch-Skea (FS). 

For the Buchdahl ansatz (\ref{eq_B}),  Eqs. (\ref{eqn_2} -- \ref{eqn_4}) take the following form 
 
\begin{eqnarray}
\rho (r)= \frac12\frac{(n-2)!}{(n-2N-1)!}\left(\frac{A}{1+C r^2}\right)^N\left(n-1-2N+\frac{2N}{1+C r^2}\right)
\label{eqn_6}
\end{eqnarray}

\begin{equation}
p(r)= \frac{(n-2)!}{(n-2N-1)!}\left(\frac{N(1-(A-C) r^2)}{A r}\frac{f'_2(r)}{f_2(r)}-\frac12 (n-2N-1)\right) \left(\frac{A}{1+C r^2}\right)^N
\label{eqn_7}
\end{equation}

that can also be written as:

\begin{equation}
p(r)=\rho\left(-1+ \frac{2N}{Cr^2(n-2N-1)+n-1}\left(1+\frac{(1+Cr^2)(1-(A-C) r^2)}{A r}\frac{f'_2(r)}{f_2(r)}\right)\right)
\label{eqn_7bis}
\end{equation}

\noindent
and the pressure isotropy equation takes the form 
\begin{eqnarray}
&&
(1-(A-C) r^2)(1+C r^2)f''_2(r)
+\frac{f'_2(r)}{r}((2N-1)C(A-C) r^4-2NC r^2-1)+\nonumber\\&& AC (n-2N-1) r^2 f_2(r)=0. 
\label{eqn_8}
\end{eqnarray} 

As in Ref. \cite{Buchdahl} we write $r$ to $x=r^2$ to cast the above equation in the form 
\begin{eqnarray}
&&(1-(A-C) x)(1+C x)f''_2(x)
+ f'_2(x)\left((A-C)(N-1)C x-C(N-1)-\frac{A}{2}\right)+\nonumber\\&&\frac14 AC(n-2N-1) f_2(x)=0\label{eqn_8bis}
\end{eqnarray} 
where a prime here as well as henceforth will indicate derivative relative to argument.

Now there arise two cases corresponding to $A\neq C$ (BVT) and $A=C$ (FS). 

\subsection{Buchdahl-Vaidya-Tikekar model}

When $A>C>0$, we do the following change of variable
\begin{equation}
 z=\frac{A-C}{A}(1+C x)\label{eqn_var1}
 \end{equation}
then the isotropy equation becomes 
\begin{equation}
z(1-z)F''(z)+((N-1)z+1/2-N)F'(z)+\frac14\frac{A}{A-C}(n-2N-1)F(z)=0.
\label{eqn_9}
\end{equation}
This is the Gauss equation \cite{Gauss}
$$(1-z)zF''(z)+[c-(a+b+1)z]F'(z)-abF(z)=0$$
with $c=1/2-N$, $a+b=-N$, $-ab=A (n-2N-1)/(4(A-C))$. The equation in question can be easily solved and the two independent solutions of equation (\ref{eqn_9}) around $z=0$ for $n>2N+1$ are the hypergeometric functions, $_2F_1(a,b;c,z)$ and $z^{1-c}\,{}_2F_1(1+a-c,1+b-c;2-c,z)$.

As is the case for N=1 Einstein gravity \cite{kdm}, the equation (\ref{eqn_9}) is the same  for a given Lovelock order $N$ in all $n>2N+1$ dimensions with the constants $A$ and $C$ are related as follows: 
$$\frac{A_n}{A_n-C_n}(n-2N-1)=\frac{A_{2N+2}}{A_{2N+2}-C_{2N+2}}$$
Here a subscript refers to spacetime dimension. It means that a $(2N+2)$-dimensional solution could be lifted to a higher $n$-dimensional solution with $K_{2N+2}$ being replaced by $K_n$ according to the above relation. This is because for a given $N$, the equation has the same form and hence the same solution with appropriate $K$ parameter. 

For the VT case, this relation takes the form  
\begin{equation}
K_n=\frac{K_{2N+2}+2N+2-n}{n-2N-1}\label{eqn_GR}
\end{equation}
This is the pure Lovelock generalization of the Einstein gravity relation for $N=1$ obtained in the paper I \cite{kdm}.

The general solution of Eq. (\ref{eqn_9}) around $z=0$ is  given by
\begin{equation}
F^B(z)=A_1 F^B_1(z)+A_2 F^B_2(z)
\label{eqn_10}
\end{equation}
where 
\begin{equation}
F^B_1(z)\equiv {_{2}F_{1}}(a,b;1/2-N,z),\quad F^B_2(z)\equiv z^{N+1/2}{_{2}F_{1}}(1/2+N+a,1/2+N+b;3/2+N,z)
\label{eqn_10bis}
\end{equation}
Here $A_1$ and $A_2$ are arbitrary integration constants and ${_{2}F_{1}}(a,b;c,z)$ is the hypergeometric function with 
$$a=-\frac{N}{2}+\frac12\sqrt{N^2+\frac{A}{A-C}(n-2N-1)},\quad b=-\frac{N}{2}-\frac12\sqrt{N^2+\frac{A}{A-C}(n-2N-1)}.$$
Note that $z=(A-C)(1+Cr^2)/A$ and when the expression under the radical is whole square, the hypergeometric function becomes a polynomial. We thus have the complete solution for the isotropy equation for the Buchdahl or Vaidya-Tikekar ansatz.

\subsection{Finch-Skea model}

When $A=C$ we write  for $n>2N+1$  
\begin{equation}
 z=(1+C x)(n-2N-1)\label{eqn_var2}
 \end{equation}
and the isotropy equation (\ref{eqn_8bis}) becomes
\begin{equation}
zF''(z)-\left(N-\frac{1}{2}\right)F'(z)+\frac14  F(z)=0. 
\label{eqFS1}
\end{equation}
where now $z$ is given in (\ref{eqn_var2}) and    
 \begin{equation}
 F^{FS}_1(z)\equiv z^{(N+1/2)/2} J_{N+1/2}(\sqrt{z}),\quad F^{FS}_2(z)\equiv z^{(N+1/2)/2} J_{-N-1/2}(\sqrt{z}) \label{eqn_FS3}
 \end{equation} 
 where $J$ are the Bessel functions. The solution for a given Lovelock order $N$ is the same for any dimension $n>2N+1$.

This is the Bessel equation and it is important to note it remains the same for a given $N$ in all dimensions. That is, for a given Lovelock order $N$, the solution is universal for the variable $z$ in all dimensions $\geq2N+2$. The general solution is given by 
\begin{equation}
 F^{FS}(z)=A_1 F^{FS}_1(z)+A_2 F^{FS}_2(z)
 \label{eqn_FS2}
 \end{equation}

Though in the critical odd $n=2N+1$ dimensions there can be no bound fluid distributions, yet the solution of the isotropy equation (\ref{eqn_8}) for BVT and FS ansatz are respectively given as follows:   
\begin{equation}
F^B_{n=2N+1}(z)= A_1+A_2\,\,{_{2}F_{1}}(1/2+N,1/2;3/2+N,z)
\end{equation}
where $z$ is given in (\ref{eqn_var1})
and
\begin{equation}
  F^{FS}_{2N+1}(z) =  A_1+A_2 z^{N+1/2}\label{eqn_FS4}.
\end{equation}  
where now $z=1+C r^2$.

\section{Matching with the exterior solution} 

At the star boundary which is defined by $p=0$, the solution must match with the pure Lovelock vacuum solution in the exterior. This requires that the metric functions $g_{tt}, g_{rr}$ and $g'_{tt}$ must be continuous across the boundary. 

In the interior for $n\geq 2N+2$, we have  
\begin{equation}
ds^2=-f_2(r)^2 dt^2+\frac{1+Cr^2}{1+(C-A)r^2}dr^{2}+ r^{2}d\Omega^{2}_{n-2},
\label{eqn_20int}
\end{equation}
where $f_2(r)$ is $F^B(z_B)$ and $F^{FS}(z_{FS})$ respectively for Buchdahl and FS models and
\begin{equation}
z_B=\frac{(A-C)(1+C r^2)}{A},\quad z_{FS}=(1+Cr^2)(n-2N-1).
\label{eqn_32}
\end{equation}
Since It has to be matched to the pure Lovelock vacuum exterior metric \cite{dpp}
\begin{equation}
ds^2=-f(r) dt^2+f^{-1}(r)dr^{2}+ r^{2}d\Omega^{2}_{n-2},
\label{eqn_20ext}
\end{equation}
where
\begin{equation}
f(r)=1-2Mr^{-\frac{n-2N-1}{N}}.
\label{eqn_21}
\end{equation}
The continuity of the metric components determines mass 
\begin{equation}
M=\frac12\frac{A r_0^{(n-1)/N}}{1+C r^2_0},
\label{eqn_22}
\end{equation}
where we have
$A=(K+1)\alpha,\, C=K\alpha$ for VT and $A=C$ for FS, and 
in  both cases $g_{tt}$ can be written as $g_{tt}=-(A_1 g_1(r)+A_2g_2(r))^2$. Taking continuity of $g_{tt}$ and  $g'_{tt}$ and some further manipulations lead to 
\begin{equation}
A_1 g_1(r_0)+A_2 g_2(r_0)=\sqrt{1-\frac{Ar_0^2}{1+Cr_0^2}}
\end{equation}
\begin{equation}
A_1 g_1'(r)|_{r=r_0}+A_2 g_2'(r)|_{r=r_0}=\frac{r_0 A(n-2N-1)}{2N\sqrt{(1+Cr_0^2)(1+(C-A)r_0^2)}}
\end{equation}
and then $A_1$ and $A_2$ are determined as 
\begin{equation}
A_1=\left.\frac{\beta \, g_2'(r)-\delta \, g_2(r)}{g_1(r)g_2'(r)-g_2(r)g_1'(r)}\right|_{r=r_0},\qquad
A_2=-\left.\frac{\beta \, g_1'(r)-\delta \, g_1(r)}{g_1(r) g_2'(r)-g_2(r)g_1'(r)}\right|_{r=r_0} ,
\label{eqn_23}
\end{equation}
where
\begin{equation}
\beta=\sqrt{\frac{1-(A-C) r^2_0}{1+C r^2_0}} ,\quad
\delta=\frac{A r_0(n-2N-1)}{2N\sqrt{(1+C r^2_0)(1+(C-A)r_0^2)}}
 \label{eqn_25}
\end{equation}
where $r_0$ is the star radius. For  $0\leq r \leq r_{0}$, the pressure from Eq. (\ref{eqn_7bis}) is determined as 
\begin{equation}
p(r)=\rho\left(-1+ \frac{2N}{Cr^2(n-2N-1)+n-1}\left(1+\frac{(1+Cr^2)(1-(A-C) r^2)}{A r}\frac{f_2'(r)}{f_2(r)}\right)\right)\label{eqn_31}
\end{equation}
where $f_2(r)$ is as given in Eqs. (\ref{eqn_10}--\ref{eqn_10bis}) and (\ref{eqn_FS3}--\ref{eqn_FS2})   for Buchdahl and Finch Skea models respectively.

This completes the matching. 

\section{Physical properties}

One of the most important properties for a star model is the compactness. For a given radius how much mass could be packed in. In Ref. \cite{Buchdahl}, Buchdahl has also obtained the compactness limit given by 
\begin{equation}
r_0>9M/4
\end{equation}
by requiring density to decrease monotonically from the center. This limit also follows from the constant density distribution \cite{dad-compact} for which sound speed becomes infinite. This naturally defines the limiting compactness. 

From Eq. (32), it is clear that $M(C= 0) > M(A>C) > M(A=C)$ indicating for a given star radius, mass is maximum for constant density distribution ($C=0$ and minimum for FS ($A=C$ and BVT ($A>C$) lies in between these two limiting distributions. This equation also suggests that volume of star goes as $r_0^{(n-1)/N)}$ for an effective density, $A/2(1+Cr_0^2)$. Now for the critical even dimension $n=2N+2$, we have volume going as $r_0^{2+1/N}$ collapsing to $r_0^2$ for $N\to\infty$. It seems to indicate that in the large $N$ limit volume effectively becomes area! This clear shows that volume and thereby mass is maximum for $N=1$ and it goes on decreasing with $N$. This is because unlike Einstein gravity, \footnote{Gravitational potential  for Einstein goes as $1/r^{n-3}$ while for pure Lovelock as $1/r^{(n-2N-1)/N}$.} pure Lovelock gravity becomes weaker with increasing $N$. This means stars are densest for Einstein gravity and they become rarer as $N$ increases.

On the hand if we look at density as given in Eq. (12 ), at the center it goes as $A^N$ for Buchdahl and $C^N$ for FS, and $A > C$ always. For a given star radius $r_0$, constant density $\rho_{const}$ will be the maximum, and hence $\rho(r=0)\leq \rho_{const}$. Note that constant density cannot be reached for FS because it requires $C=0$. \footnote{This is because the two define the extremity limits and hence they both have to be exclusive.} Since $A\geq C$, hence we shall have  $\rho_{FS}(r=0)\leq \rho_{B}(r=0)\leq \rho (const.)$. This clearly indicates the degree of compactness with constant density giving the upper limit while FS the lower limit, and all other physically acceptable models lying in between.    

As an aside we also give pressure for constant density star with $K=0$ in VT model, which means $C=0$ for Buchdahl, and it is given by 
\begin{equation}
p = \frac{n-2N-1}{n-1}\left(-1+\frac{2N}{n-1}\frac{1}{\displaystyle{\frac{\sqrt{1-A r_0^2}}{\sqrt{1-A r^2}}-\frac{n-2N-1}{n-1}}}\right)
\end{equation}
for any Lovelock order $N$. 

\subsection{Density and pressure}
 
For $K=0$ we have the constant density solution  $$g_{rr}=\frac{1}{1-A r^2,}$$  with $C=0$ and if we have $T^t_t=-\rho$ where
$\rho=constant$
$$ A=\left(2\rho \frac{(n-2N-1)!}{(n-1)!}\right)^{1/N}.$$
In the plots we have taken the above value for $A$ and have set $\rho=1$, and have set $n=2N+2$, the critical dimension.

We are going to plot the normalized density and pressure $\rho/\rho(r=0), p/p(r=0)$ for VT and FS models for Einstein and GB gravity. We take $K=7, 14$ and $N=1, 2$ for $n=4, 6$. It turns out that normalized density has the same behavior for both VT and FS (Fig. 1) while the pressure plots (Figs 2, 3) differ. 

Also note the constants $A$ and $C$ for $n=2N+2$ are given by
$$C=K\left(\frac{2}{(2N+1)!})^{1/N}\right), \quad A=(K+1)\left(\frac{2}{(2N+1)!})^{1/N}\right) .$$

\begin{figure}
\includegraphics[width=7cm]{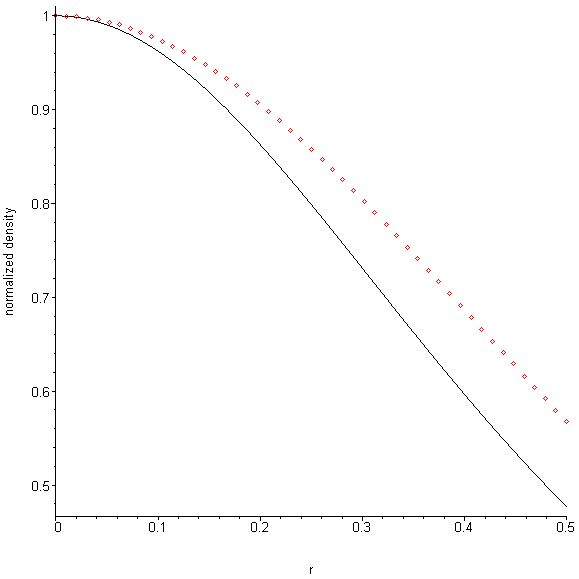}\hspace*{1cm}\includegraphics[width=7cm]{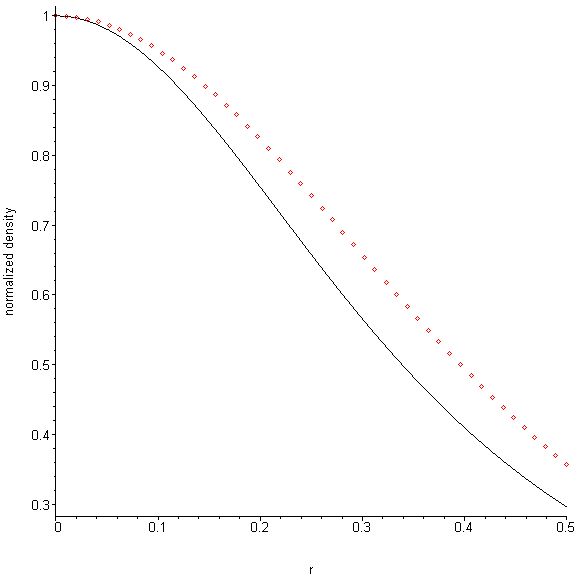}
\caption{\label{fig:densityN12} Normalized density plots for both VT and FS in the critical dimensions $2N+2$ for $N=1, n=4$ Einstein (black) and $N=2, n=6$ GB(red). On the left is $K=7$  and $K=14$ on the right.}
\end{figure}

\begin{figure}
\includegraphics[width=7cm]{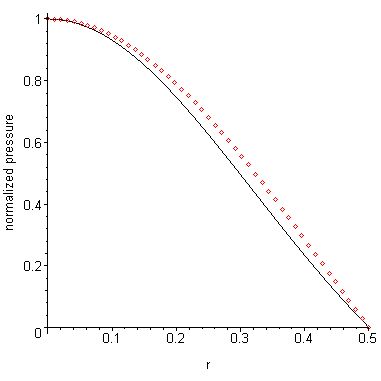}\hspace*{1cm}
\includegraphics[width=7cm]{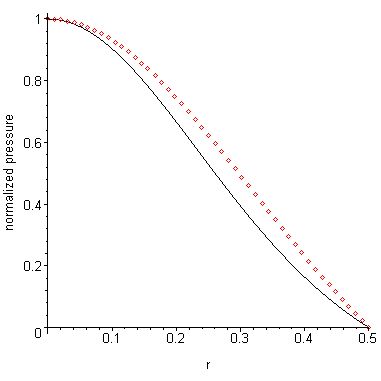}
\caption{\label{fig:pressureBN12} Normalized pressure plots for VT in the critical dimensions $2N+2$ for $N=1, n=4$ Einstein (black) and $N=2, n=6$ GB(red). On the left is $K=7$  and $K=14$ on the right.} 
\end{figure}

\begin{figure} 
\includegraphics[width=7cm]{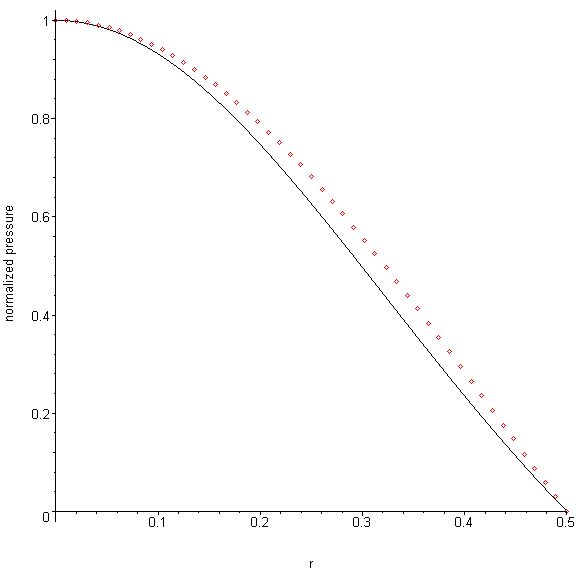}\hspace*{1cm}
\includegraphics[width=7cm]{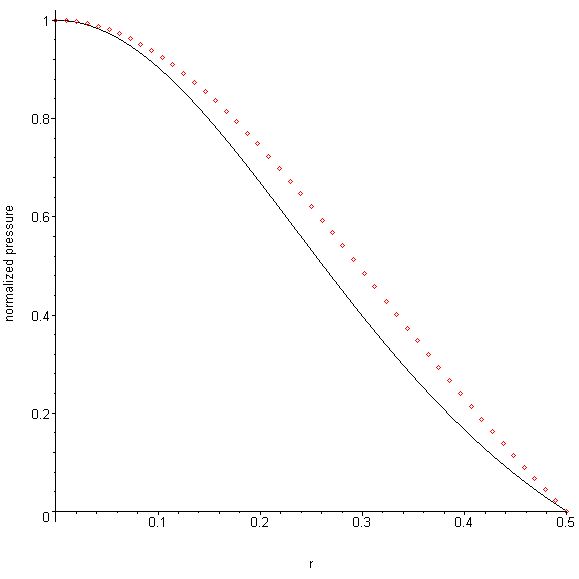}
\caption{\label{fig:pressureFSN12} Normalized pressure plots for FS in the critical dimensions $2N+2$ for $N=1, n=4$ Einstein (black) and $N=2, n=6$ GB(red). On the left is $K=7$  and $K=14$ on the right.} 
\end{figure}

This may however be noted that though the normalized plots for pressure look quite similar but the actual value differ quite significantly in various cases. In the Table 1 we give some representative values of pressure at $r=0$.

\begin{table}
\begin{tabular}{|l|r|r|r|r|r|r|}
\hline
$N$, $n$ $K$ & 1, 4, 7 & 1, 4, 14 &  2, 6, 7 & 2, 6, 14 & 2, 7, 3/2 & 2, 7, 5 \\ \hline
$p(r=0)$ FS & 0.9109   & 3.6264   & 1.1276   & 7.6555 & 0.0154 & 0.5413\\
$p(r=0)$ B  & 1.3109   & 4.6712   & 1.7299   & 9.6678 & 0.0725 & 0.9531\\ \hline
\end{tabular}
\caption{The first row in the table indicates $N, n, K$ while the second and the third give central pressure values for FS and B.}
\end{table}

For VT models, recall the relation $K_n=(K_{2N+2}-n+2N+2)/(n-2N-1)$. Let us consider $N=2$ GB case with $K_6=4$ giving $K_n=4, 3/2, 2/3, 1/4, 0$ for $n=6, 7, 8, 9, 10$, and similarly for $K_6=11$, $K_n=11, 5, ..., 0$ for  $n=6, 7, ..., 17$. The normalized density and pressure (normalization done relative to the central value) for these cases are plotted in Figs. 4, 5. As dimension increases density goes on increasing until it reaches constant density corresponding to $K=0$ determining the maximum dimension for a given initial $K_{2N+2}$. The pressure however does not show much marked difference between various cases.

\begin{figure}
\includegraphics[width=7cm]{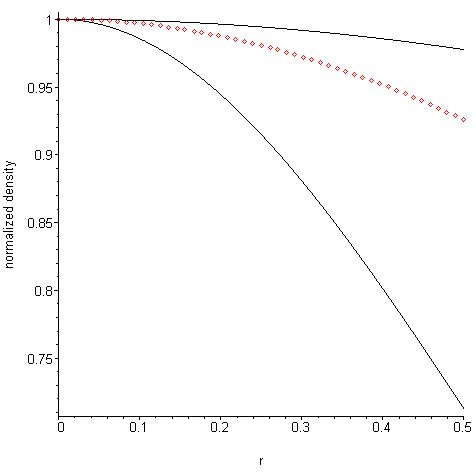}\hspace*{1cm}
\includegraphics[width=7cm]{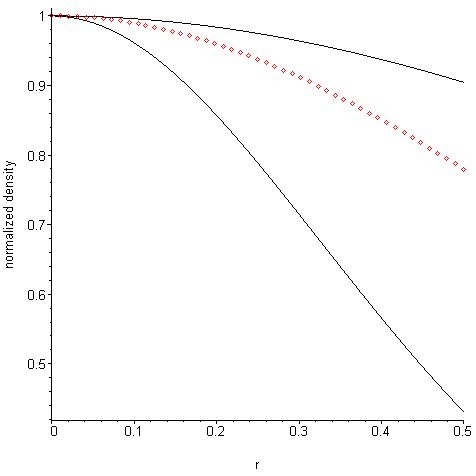}
\caption{\label{fig:densityN2} 
The normalized density, $\rho(r)/\rho(r=0)$, plots show in ascending order for the cases: $K_6=4$ and $n=6, 7, 8$ on the left and $K_6=11$ and $n=6, 7, 8$ on the right respectively.}
\end{figure}

\begin{figure}
\includegraphics[width=7cm]{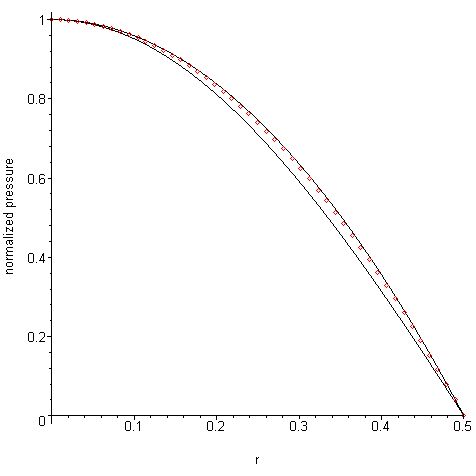}\hspace*{1cm} \includegraphics[width=7cm]{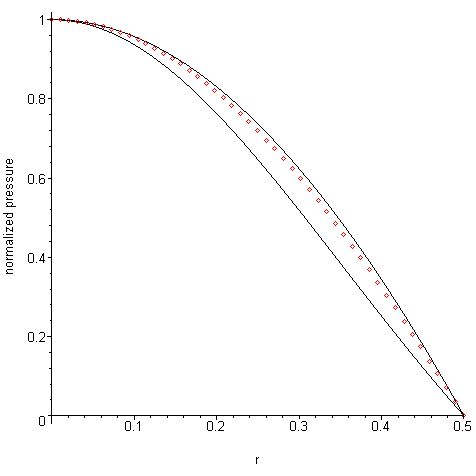}
\caption{\label{fig:pressureN2} 
The normalized pressure, $p(r)/p(r=0)$, plots show in ascending order for the cases: $K_6=4$ and $n=6, 7, 8$ on the left and $K_6=11$ and $n=6, 7, 8$ on the right respectively.}
\end{figure}

\subsection{Sound speed}
The  sound speed in a fluid is defined as 
$$c^2_s\equiv\frac{dp}{d\rho}=\frac{p'(z)}{\rho'(z)}$$ 
From the expression for pressure in Eq. (14), we have  $p=\rho(-1+g(r))$ and so we write 
$$\frac{p'}{\rho'}=-1+g(r)+\frac{\rho}{\rho'}g'(r)$$
and  
\begin{equation}
\frac{\rho}{\rho'} = -\frac{(1+Cr^2)((1+Cr^2)(n-2N-1)+2N)}{2CNr((1+Cr^2)(n-2N-1)+2N+2)}
\end{equation}
After some algebra and using the isotropy equation for $f_2''(r)$ we obtain 
\begin{eqnarray}
S\equiv c^2_s=\frac{1+Cr^2}{Cr(Cr^2(n-2N-1)+n+1)}\frac{f_2'(r)}{f_2(r)} \left( 1+\frac{(1+Cr^2)(1+(C-A)r^2)}{Ar}\frac{f_2'(r)}{f_2(r)}\right)
\end{eqnarray}

\begin{figure}
\includegraphics[width=7cm]{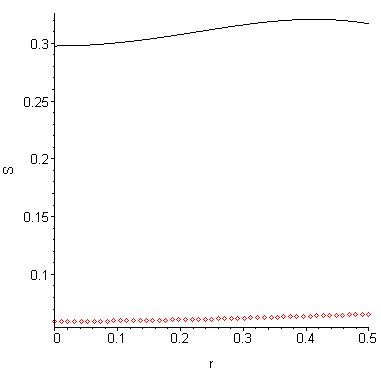}\hspace*{1cm}
\includegraphics[width=7cm]{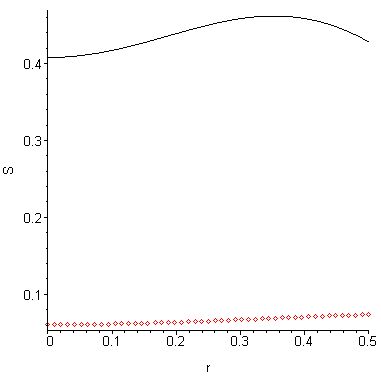}
\caption{\label{fig:soundspeedBN12} Square of sound speed plots for BVT corresponding to $N=1$ in black and $N=2$ in red for $K=7$ on the left and for $K=14$ on the right.}
\end{figure}

\begin{figure} 
\includegraphics[width=7cm]{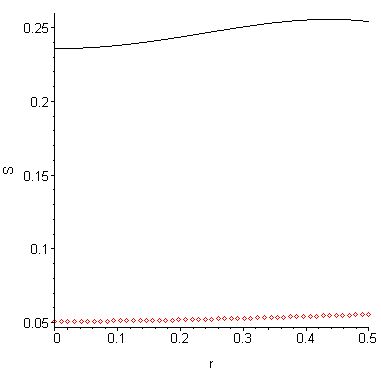}\hspace*{1cm}
\includegraphics[width=7cm]{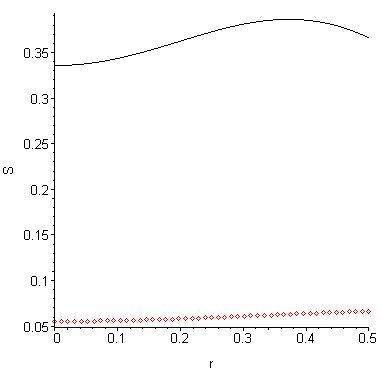}
\caption{\label{fig:sounspeedFSN12} Square of sound speed plots for FS corresponding to $N=1$ in black and $N=2$ in red for $K=7$ on the left and for $K=14$ on the right.}
\end{figure}

In this expression we must use for $f_2(r)$ and its derivative the corresponding results obtained after the matching with the exterior solution. The plots for the sound speed, $v=c_s^2$ using the same values of the parameters as for the density and pressure plots and  the constants $C$ and $A$ as given in in Eqs. (\ref{fig:soundspeedBN12}) and (\ref{fig:sounspeedFSN12}). For FS we have $A=C$. 

In Figs 6 and 7 we plot square of sound speed, $v=c_s^2$ for VT and FS (For FS, we use the same constant $C$) for $N=1$ in black and $N=2$ in red for $K=7$ on the left and for $K=14$ on the right. Clearly sound speed is much lower for GB as compared to Einstein case. This would be the trend for pure Lovelock as $N$ increases sound speed will go on decreasing. This is because gravitational potential going as $1/r^{(n-2N-1)/N}$ becomes weaker and weaker as $N$ increases. In other words, distribution becomes less compact with increasing $N$. 
 
\section{Discussion} 

In the paper I we had considered the general Buchdahl ansatz \cite{Buchdahl} covering all  
the physically interesting star models for Einstein gravity and had shown that how a $4$-dimensional solution could be taken over to higher dimensions by properly redefining the Vaidya-Tikekar parameter $K$ marking deviation from sphericity of $3$-space geometry. It is remarkable that this geometrical property has the physical imprint in the requirement that  $K$ has to be non-negative for density to monotonically decrease outwards. This happens because the only equation to be solved is that of the pressure isotropy which remains in the Gauss form in all dimensions.  

In this paper we extend this framework from Einstein to pure Lovelock gravity and show that the same features persist. It turns out that for a given $N$, the equation has the same Gauss form for BVT while for FS it is the Bessel equation in all dimensions $\geq 4$. Thus fluid solutions for a star interior have universal character relative to a suitably defined variable and the $K$ parameter for VT models. This universal behavior is true only for pure Lovelock gravity and hence this is yet another of its distinguishing features \cite{dad2, sum-dad}.

One of the most pertinent questions for a star model is its compactness. The constant density distribution is obviously the most compact with star radius, $r_0>9M/4$, the Buchdahl limit \cite{Buchdahl} \footnote{It has recently been generalized \cite{dad-compact} for pure Lovelock to read as $r_0^\alpha > 9M^{1/N}/4$ where $\alpha =(n-2N-1)/N$.}. It turns out that the other end of lower bound is defined by the FS model, and all other physically acceptable models lie between these two limiting  distributions. This is reflected clearly in mass and density spectrum as $M_{FS}(A=C) < M_{B}(A>C) < M_{\rho const.}(C=0)$ and  $\rho_{FS}(r=0)\leq \rho_{B}(r=0)\leq \rho (const.)$. Also note that FS model does not admit constant density distribution, this is because the two limiting cases must be exclusive. This is indeed a very important and interesting property which, so far as we know, has not been earlier reported in the literature. 

As $N$ increases gravitational potential for pure Lovelock gravity becomes weaker and hence  consequently star interior becomes rarer; i.e. for a given radius the most compact distribution would be for $N=1$ Einstein and it goes on becoming less compact with increasing $N$. What happens is that volume in the critical dimension $n=2N+2$ goes as  $r_0^{2+1/N}$ which collapses to two dimensional volume -- area in the limit $N\to\infty$. It indicates that in large $N$ limit space dimension seems to collapse to two! It clearly shows volume and thereby mass for a given star radius is maximum for $N=1$ Einstein gravity, and hence it is most compact. 

We have found the general equations for the density, pressure and pressure isotropy for a spherical symmetric perfect fluid in any pure Lovelock order and in any dimension bigger than $2N+1$.

We have shown that $4$-dimensional solution could be taken over to higher dimensions in both Einstein (paper I) as well as in pure Lovelock theory, the important question that arises is that would the higher dimensional solution be stable? This is what we would like to study in the future investigation. 

\section*{Acknowledgements}
AK and AM gratefully acknowledge IUCAA for the invitation and warm hospitality which facilitated this collaboration. Partial support for this work to AK was provided by Uzbekistan Foundation for Fundamental Research project F2-FA-F-116. Partial support for this work to AM was provided by FIS2015-65140-P (MINECO/FEDER).  ND thanks Albert Einstein Institute, Golm and the University of Barcelona for visits that facilitated finalization of the manuscript.


\begin{thebibliography}{99}
\bibitem{sum-dad} S Chakraborty, N Dadhich, Do we really live in four or higher dimensions; arxiv:1605.01961
\end{thebibliography}
\end{document}